\documentclass[12pt]{article}

\usepackage{psfig}

\begin{document}

% Some macros for this paper

%defsnew
%\def\i{\infty}  %overwrites plain tex defn    
\def\L{\Lambda} %overwrites plain tex defn 
\def\l{\lambda} %overwrites plain tex defn  
\def\v{\n\ll}   %overwrites plain tex defn
\def\t{\tau}    %overwrites plain tex defn
\def\o{\over}   %overwrites plain tex defn
\def\b{\beta}   %overwrites plain tex defn

\def\f{\phi}         \def\p{\partial}
\def\G{\Gamma}   \def\k{\kappa}  
\def\ra{\rightarrow}  \def\kl{\kappa L}       \def\lb{\bar\l}  
\def\ft{\vf^2}        \def\klr{\kl\rho}    \def\fb{\bar\f_B}     
\def\cf{\cal F}
\def\e{\varepsilon}  \def\GN{\Gamma^{(N)}}  
\def\gf{\gamma_{\pi}}  \def\tb{\bar t}      \def\fbd{\fb'}
\def\ff{\phi^4}  
\def\vf{\varphi}
\def\r{\rho} 
\def\gft{\gamma_{\vf^2}}  \def\ppk{\k{\p\over{\p\k}}}
  
\def\s{\scriptstyle}     \def\ss{\scriptscriptstyle}
\def\disp{\displaystyle}           
\def\kc{\k_c}
\def\y{L\o\xi_L}         \def\rc{\r_c}    
\def\Gc{\cal G}    \def\F{\cal F}      \def\M{\cal M}
\def\T{\cal T}
\def\g{\gamma}
\def\gl{\gamma_{\l}}
\def\glb{\gamma_{\lb}}

\def\a{\alpha}
\def\cG{{\cal G}}
\def\hf{\hat\Phi}
\def\cG{{\cal G}}
\def\th{\theta}
\def\A{\cal A}
\def\dea{\de^{\ast}}
\def\ef{_{\ss eff}}
%%%%%%%%%%%%%%%%%%%%%%%%%%%%%%%%%%%%%%%%%%%%%%%%%%%%%%%%%%%%%%%%
%%additional definitions:
\def\Got{\G^{(0,2)}}
\def\Gnl{\G^{(N,L)}}
\def\R{\rlap I\mkern3mu{\rm R}} 
\def\be{\begin{equation}}
\def\ee{\end{equation}}

%%%%%%%%%%%%%%%%%%%%%%%%%%%%%%%%%%%%%%%%%%%%%%%%%%%%%%%%%%%%%%%
\begin{flushright}
%\today
\end{flushright}
\vskip 22truept
\begin{center}
{\LARGE Dimensional Crossover in the Non-Linear Sigma Model}
\vskip\baselineskip
\vskip 15pt
{\bf Denjoe O' Connor},\\
\vskip 15pt
Departmento de F\'\i sica, Cinvestav, \\
Apartado Postal 14-740,\\ 
M\'exico D.F. 07000, M\'exico.\\
Denjoe.Oconnor@fis.cinvestav.mx
\vskip\baselineskip
\vskip 15pt
{\bf C.R. Stephens},\\
\vskip 15pt
Instituto de Ciencias Nucleares, U.N.A.M,\\
Circuito Exterior, Apartado Postal 70-543\\
M\'exico D.F. 04510, M\'exico.\\
stephens@nuclecu.unam.mx
\vskip\baselineskip
\vskip 15pt
{\bf Jos\'e Antonio Santiago},\\
\vskip 15pt
Instituto de Ciencias Nucleares, U.N.A.M,\\
Circuito Exterior, Apartado Postal 70-543\\
M\'exico D.F. 04510, M\'exico.\\
santiago@nuclecu.unam.mx
\vskip 0.1truein
PACS numbers: 64.60.Ak, 05.70.Fh, 11.10.Gh\\
\end{center}

\vskip 0.3truein{\bf Abstract:} 

We consider dimensional crossover for an $O(N)$ model on a 
$d$-dimensional layered geometry
of thickness $L$, in the $\sigma$-model limit, using
``environmentally friendly'' renormalization. We show
how to derive critical temperature shifts, giving explicit 
results to one loop. We also obtain expressions for the 
effective critical exponents $\delta_{\rm eff}$ and 
$\beta_{\rm eff}$ that interpolate between their characteristic
fixed point values associated with a $d$ and $(d-1)$-dimensional
system in the limits $T\rightarrow T_c(L)$, with 
$L(T-T_c(L))^{\nu}\rightarrow\infty$, and $T\rightarrow T_c(L)$
for $L$ fixed respectively, where $T_c(L)$ is the $L$-dependent
critical temperature of the system.   

\vfil \eject

\section{Introduction}

Crossover behavior --- the interpolation between qualitatively
different effective degrees of freedom of a system as a function of
scale --- is both ubiquitous and extremely important. Calculation of
scaling functions associated with crossover behavior is, generally
speaking, much more difficult than the calculation of more standard
universal quantities, such as critical exponents, the latter being
calculable in an approximation scheme suitable for the asymptotic
region around one critical point.

An important, non-trivial and experimentally accessible example is
seen in the context of  confined systems and their analysis via finite
size scaling. As far as the fluctuations in a system are concerned
there is, in principle, a very marked difference between an
``environment'' consisting of infinite three dimensional space and a
three dimensional box of  ``size'' $L$.  A general formalism for
studying such crossover systems using renormalization group (RG)
methods is that of ``environmentally friendly'' renormalization
\cite{us}. To access the sensitivity to environment implicit in such
a system it is necessary to implement a renormalization programme
which is explicitly dependent on the relevant environmental
parameters, such as finite size $L$. By so doing one can access
several fixed points of one globally defined RG, and what is more, one
may achieve this perturbatively using one uniform approximation scheme.

The main gist of the approach is based on the simple intuition that,
viewed as a coarse graining procedure, a ``good'' coarse graining will
be one that when effected to a length scale comparable to any length
scale set by the  environment will reflect the influence of the latter
by changing continuously as a  function of scale the type of effective
degree of freedom being coarse grained.  However, and this is a point
to be emphasized, although the intuition is grounded in a coarse
graining procedure the actual mechanics are totally different to that
of a Kadanoff/Wilson type coarse graining.  Instead the formalism is
based on the  notion of a RG as describing the invariance under
reparametrization of a system, an idea which goes back to the original
formulation of the field theoretic RG back in the '50's. Just as
there are good and bad coarse grainings so there are good and bad
reparametrizations. An ``environmentally friendly'' reparametrization is
one that tracks the qualitatively changing nature of the effective
degrees of freedom of a  crossover system. As has been emphasized
previously a necessary  condition for an enviromentally friendly RG to
satisfy is that the number of fixed points of the RG, defined globally
on the space of parameters, be diffeomorphic to the number of points
of scale invariance of the system.  Unfortunately, favourite forms of
field theoretic renormalization, such as minimal  subtraction,
manifestly do not satisfy this criterion and therefore are of only
limited use in describing crossover behavior.

Previously \cite{us}, we have considered crossover behaviour for an 
$O(N)$ model in the context of a Landau-Ginzburg-Wilson representation of the 
underlying lattice model based on a $\lambda\vf^4$ theory. Given that
for a dimensional crossover $\varepsilon$ expansion methods cannot work
a fixed dimension expansion was used to access the crossover the essential
characteristics of the perturbation theory being based on an expansion around
the Gaussian fixed point. As is well
known, when an $O(N)$ symmetry is spontaneously broken massless Goldstone
modes give singularities at large distances for any value of the temperature.
The thermodynamics of these spin waves is described in the long distance
limit by a Landau-Ginzburg-Wilson Hamiltonian which is that of the field
theoretic non-linear $\sigma$-model \cite{sigma}. The appropriate 
expansion parameter in this case is the temperature, $T$, and hence 
perturbation theory corresponds to a low temperature expansion. 

Crossover behaviour in the context of the non-linear $\sigma$-model has
been considered previously. In particular, Amit and Goldschmidt extended
their original treatment of bicritical systems above the critical temperature
using Generalized Minimal Subtraction \cite{AG} to below the critical
temperature \cite{AGsigma} thus describing the crossover between a 
system exhibiting an $O(N)$ symmetry to that of an $O(M)$ symmetry. In this
case $\varepsilon$ methods were perfectly feasible due to the fact
that the upper critical dimension of the two fixed points was the same.
In the context of finite size scaling Br\'ezin and Zinn-Justin described
crossover of the non-linear $\sigma$-model in the context of a box or a 
cylinder \cite{brezzj} by treating the lowest infra-red modes of the 
system non-perturbatively while treating other modes in a perturbative
$\varepsilon$ expansion. However, this method does not work in the context
of a dimensional crossover such as in a thin film where the reduced
dimension system also exhibits a non-trivial fixed point. 
 
More recently, the quantum version of the non-linear $\sigma$-model 
\cite{chak} has generated a great deal of attention as it can 
describe the long-wavelength, low-temperature behaviour of a 
two-dimensional quantum Heisenberg ferromagnet which in turn has
been proposed as a model of high-temperature superconductors. Given the
close analogy between a $(d+1)$-dimensional layered system with periodic 
boundary conditions and a $d$-dimensional quantum system it is certainly
of interest from this point of view to investigate further dimensional
crossover in the context of the non-linear $\sigma$-model.

In this paper we will consider dimensional crossover of a ferromagnet
with $O(N)$ symmetry in the broken phase using as starting point 
the non-linear $\sigma$-model and utilizing the techniques of 
environmentally friendly renormalization to access the full universal 
crossover behaviour.  One of the benefits of doing so is a better understanding
of how environmentally friendly renormalization functions
in the context of a low temperature expansion as opposed to an
expansion around the critical point.

In section 1 we briefly outline some important features of the
non-linear $\sigma$-model. In section 2 we consider some formal
renormalization results leaving explicit one loop answers for the
case of dimensional crossover in a film geometry to section 3. In
section 4 we calculate the shift in critical temperature due to finite
size effects while in section 5 we derive one-loop expressions for
some relevant effective critical exponents in the broken phase. 
Finally, in section 6 we draw some conclusions.

\bigskip
\section{The Non-Linear $\sigma$-Model.}

We begin with the Landau-Ginzburg-Wilson Hamiltonian for a Heisenberg 
model with $O(N)$ symmetry in the $\sigma$-model limit on a $d$-dimensional
($d<4$) film geometry of thickness $L$ 
\be
{\cal H}[\vf_{B}]={1\o T_B}\int_0^L\int 
d^{d}x\left({1\over 2}\nabla_{\mu}{\bf\vf}_{B}^i\nabla_{\mu}
{\bf\vf}^i_{B}-
{H}^i_B(x){\bf\vf}^i_B\right).\label{ham}
\ee 
where $i\in [1,N]$, $\mu\in [1,d]$, $T_B$ is proportional to the 
temperature of the system and ${\vf^i}_B$ is subject to the constraint 
\be
{\vf}^i_B{\vf}^i_B=1 \label{constr}
\ee
We will restrict our attention here to the case of periodic boundary 
conditions.

The partition function $Z$ is obtained by performing the path integral
over the order parameter fields, $\vf^i_B$,  with the Hamiltonian
(\ref{ham}) subject to the constraint (\ref{constr}). Choosing
the direction of symmetry breaking to be along the $Nth$ direction we
define $\vf^N=\sigma$ and $\vf^i=\pi^i, \ \ (i\neq N)$. The constraint
implies that $\sigma(x)=\pm(1-\pi^2)^{1\o2}$. Thus the partition
function becomes
\be
Z[H,J]=\int\left[d\pi_B\over(1-{\pi_B^i}^2)^{1\over2}\right]
{\rm e}^{{-{1\o T_B}
\int d^dx\left({1\o2}(\nabla\pi_B^i)^2
+{1\o2}(\nabla(1-{\pi_B^i}^2)^{1\o2})^2-J^i_B\pi_B^i-H_B
(1-{\pi_B^i}^2)^{1\o2}\right) }}\label{part}
\ee 
Clearly this theory is highly non-polynomial. The non-trivial measure term,
which ensures the $O(N)$ invariance of the theory, can of course be
exponentiated and expanded in powers of $\pi^2$. These terms are necessary
to cancel corresponding $O(N)$ non-invariant terms that arise in perturbation
theory.  We will assume that such terms have been cancelled in the rest
of this paper and not consider them further. 
Rotations are implemented linearly in the $(N-1)$-dimensional
$\pi^i$-subspace and  non-linearly in the $\pi^i-\sigma$ directions. A
rotation by an infinitesimal $\omega^i$ induces the changes
\be
\delta\pi^i(x)=(1-{\pi^i}^2(x))^{1\o2}\omega^i
\ee
\be
\delta(1-{\pi^i}^2(x))^{1\o2}=-\omega^i\pi^i(x)
\ee
As long as $|\pi^i|<1$ the symmetry will remain broken. 
As $T\ra0$, $\sigma(x)\ra1$.

From the way in which $T$ appears in (\ref{part}) we can see that
an expansion in terms of temperature is equivalent to an expansion in
the number of loops, the only subtlety being that the measure term is
then linear in $T$ and therefore contributes to an higher order in $T$
than the other two terms. The free propagator for the $\pi$ field in
the absence of a magnetic field is
\be
 G_{\pi\pi}(k)={T_B\o k^2}\label{prop}
\ee 
The magnetic field coupled to the
$\sigma$ field acts as an IR cutoff. This can be seen  by expanding
the term $H_B(1-{\pi^i_B}^2)^{1\o2}$ in powers of $\pi$.  The
resulting two-point vertex function is
\be
\G^{(2)}_{\scriptstyle \pi}(k)={k^2+H_B\o T_B}\label{vertex}
\ee

From the form of the Hamiltonian, in terms of an expansion in $\pi$,
there are interactions of arbitrary order. However, interactions with
more than four powers of $\pi$ contribute at higher than one loop order, 
i.e. more powers of the
``small'' coupling $T$. Consequently to first order in $T$, i.e. one
loop, one need only consider the four-point interaction
\be
{1\o8T_B}\sum_{\ss{k_1 k_2 k}}(k^2+H_B)\pi^i_B(k_1)\pi^i_B(k-k_1)
\pi^j_B(-k_2)\pi^j_B(k_2-k)\label{inter}
 \ee
In this paper we will restrict our attention to $O(T)$ results and 
therefore will not consider higher order interactions any further.

\bigskip
\section{Renormalization}

In spite of the fact that the theory is non-polynomial, as is well
known \cite{sigma}, it is renormalizable using only two
renormalization constants $Z_T$ and $Z_{\pi}$ associated with  the
temperature and the field respectively. The relation between the bare
and renormalized parameters is
\be
T_B=Z_TT \ \ \ \ \ \ \pi^i_B=Z_{\pi}^{-{1\o2}}\pi\label{renparams}
\ee
To preserve the rotational invariance of the renormalized constraint 
the field $\sigma$ must renormalize in the same way 
as $\pi$. Invariance of the term $H_B\sigma_B\o T_B$ thereby yields the 
renormalization of $H$
\be
H_B=Z_TZ_{\pi}^{-{1\o2}}H
\ee
The bare and renormalized vertex functions are related via
\be
\G_{\pi}^{(N)}(k_i,T,H,L,\k)=Z_{\pi}^{N\o2}\G_{\pi B}^{(N)}(k_i,T_B,H_B,L,\L)
\label{renrel}
\ee
where $\k$ is an arbitrary renormalization scale and $\L$ an ultraviolet 
cutoff. The RG equation, which is a consequence of the $\k$ invariance of 
the bare theory, follows immediately on differentiating (\ref{renrel}) with 
respect to $\k$
\be
\left(\k{\p\o\p\k}+\beta_t {\p\o\p t}+\beta_H {\p\o\p H}-
{N\o2}\gf\right)\G_{\pi}^{(N)}(k_i,T,H,L,\k)=0\label{rgeqn}
\ee
where we have introduced a dimensionless temperature $t=T\k^{d-2}$ and 
$\gf={d\ln Z_{\pi}\o d\ln \k}$ is the anomalous dimension of the field. 
The two $\beta$-functions are
\be
\beta_t=(d-2)t-t{d\ln Z_T\o d\ln\k}
\ee
\be
\beta_H={d\ln Z_{\pi}\o d\ln\k}-{d\ln Z_T\o d\ln\k}
\ee

The two renormalization constants must be fixed by normalization
conditions. The essence of environmentally friendly renormalization 
is that in order to obtain a perturbatively well defined description of the
crossover the renormalization procedure must depend explicitly on $L$.
The normalization conditions we will use are
\be
T\G^{(2)}_{\pi}(k=0,t(\k,L\k),H(\k,L\k)=\k^2,L,\k)=\k^2\label{norm1}
\ee
\be
{\p\o\p k^2}T\G^{(2)}_{\pi}(k,t(\k,L\k),H(\k,L\k)=\k^2,L,\k)\vert_{\ss k=0}=1
\label{norm2}
\ee
Note that $T\G^{(2)}$ is just the inverse susceptibility associated with 
the $\pi$ field.

\bigskip
\section{Explicit Results}

We now proceed to examine the crossover perturbatively. To one loop
\be
\G^{(2)}_{\pi B}(k=0)={H_B\o T_B}+{(N-1)\o2}{H_B\o L}\sum_{n=-\infty}^{\infty}
\int{d^{d-1}p\o(2\pi)^{d-1}}{1\o p^2+H_B+{4\pi^2n^2\o L^2}}
\label{gtwobare}
\ee
Using the normalization conditions (\ref{norm1}) and (\ref{norm2}) one finds
\be
Z_{\pi}=1-{(N-1)\o2L\k}t \sum_{n=-\infty}^{\infty}
\int{d^{d-1}y\o(2\pi)^{d-1}}{1\o y^2+1+{4\pi^2n^2\o L^2\k^2}}
\label{rencon1}\ee
and
\be
Z_T=1-{(N-2)\o2L\k}t \sum_{n=-\infty}^{\infty}
\int{d^{d-1}y\o(2\pi)^{d-1}}{1\o y^2+1+{4\pi^2n^2\o L^2\k^2}}
\label{rencon2}
\ee
The $\beta$-function $\beta_t$ is thus given by
\be
\beta_t(t,L\k)=(d-2)t-{2(N-2)\o L\k}t^2 \sum_{n=-\infty}^{\infty}
\int{d^{d-1}y\o(2\pi)^{d-1}}{1\o (y^2+1+{4\pi^2n^2\o L^2\k^2})^2}
\label{betat}
\ee
There are three different fixed points associated with (\ref{betat}), 
a $d$-dimensional ultraviolet fixed point in the limit $L\k\ra\infty$, 
$\k\ra\infty$;
a $(d-1)$-dimensional ultraviolet fixed point in the limit $L\k\ra 0$, 
$\k\ra\infty$; and finally, a zero temperature infrared fixed point 
in the limit $\k\ra0$. However, the approach to $t=0$ depends on whether 
we consider $\k\ra0$ for fixed $L$ or $L\k\ra\infty$, $\k\ra0$. 

Because the coupling $t$ was made dimensionless with a factor $\kappa^{d-2}$,
in the limit $L\k\ra 0$, $\k\ra\infty$ one finds that 
$t\rightarrow O(L\kappa)$. However, as emphasized in \cite{us}, this does not
imply that in the dimensionally reduced limit that fluctuations are
unimportant, as loops enter with a factor $(L\kappa)^{-1}$, which diverges.
The issue is made more transparent in this limit by passing to a more
appropriate coupling $t'=t/L\kappa$. In the dimensionally reduced limit
$t'\rightarrow O(d'-2)$, where $d'=d-1$, while in the bulk limit 
$t'\rightarrow O(1/L\kappa)$. Thus $t'$ looks more natural in the dimensionally
reduced limit and $t$ in the bulk limit. Of course, these simple changes
of variables cannot affect the results for physical quantities they just help 
make more transparent what is going on. A coupling that is natural across
the entire crossover, the floating coupling, can be introduced \cite{us}. 
It is defined via $h=a_2t$ in the present case, where $a_2$ is the coefficient
of $t^2$ in equation (\ref{betat}).
 
In terms of the floating coupling one finds
\be
\k{dh\o d\k}=\e(L\k)h-h^2\label{betah}
\ee
where
\be
\e(L\k)=d-3+4 {\sum_{n=-\infty}^{\infty}
\int{d^{d-1}y\o(2\pi)^{d-1}}{{4\pi^2n^2\o L^2\k^2}\o 
(y^2+1+{4\pi^2n^2\o L^2\k^2})^3}\o
\sum_{n=-\infty}^{\infty}
\int{d^{d-1}y\o(2\pi)^{d-1}}{1\o (y^2+1+{4\pi^2n^2\o L^2\k^2})^2}}
\ee
The quantity $d_{\rm eff}=2+\e(L\kappa)$ can be interpreted as a measure 
of the effective dimension of the system 
interpolating between $d$ and $d-1$ in the limits $L\k\ra\infty$ and $L\k\ra0$
respectively, where in both cases we are considering $\k\ra\infty$, i.e. the
behaviour near the critical point. 

\begin{figure}[h]
\centerline{\psfig{figure=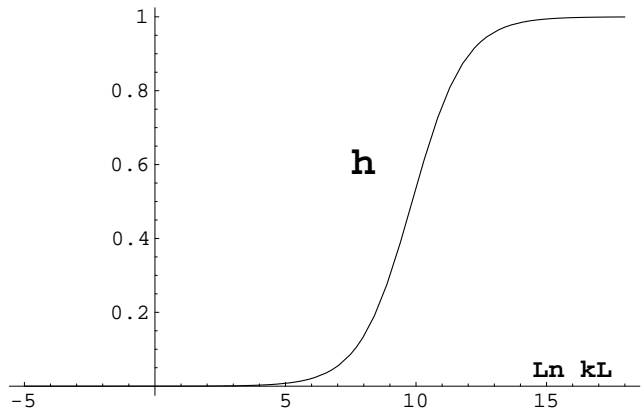,angle=-0,width=3.3in}}
\caption{Graph of separatrix solution of (\ref{betah}) as function of $\ln\kappa L$.} 
\end{figure}

The corresponding fixed points for $h$ are: $d-2$, $d-3$ and $0$.
In Figure 1 we see a plot of $h$ as a function of $\ln\kappa L$ for 
$d=3$. In this case, as the theory is asymtotically free in two 
dimensions, the coupling goes to zero in the dimensionally reduced 
limit, i.e. the $(d-1)$-dimensional ultraviolet fixed point and 
the trivial infrared fixed points coincide. This dimensional crossover
in the coupling is controllable in the low temperature expansion. 
It is in fact the 
solution to (\ref{betah}) that we use as a ``small'' parameter in the 
perturbative expansion of all other quantities. It is the fact that $h$
captures the crossover between the different fixed points that gives us a 
uniform expansion parameter and therefore perturbative control of the
crossover. Of course when $d-2$ is not small one really needs to work to
higher order and attempt some resummation method. It should be clear 
however that there is no impediment to continuing this calculation to 
arbitrary order in the loop expansion.
\begin{figure}[h]
\centerline{\psfig{figure=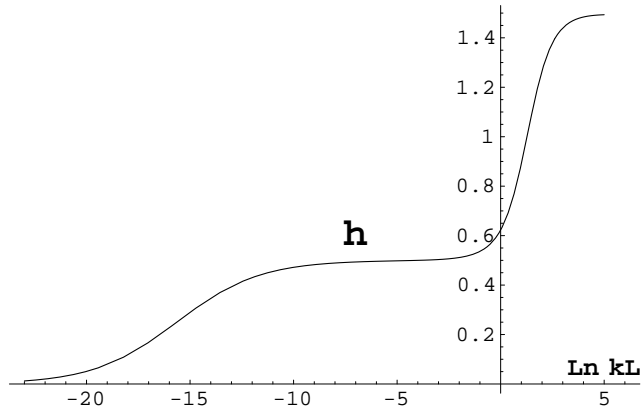,angle=-0,width=3.3in}}
\caption{Graph of separatrix solution of (\ref{betah}) as function of $\ln\kappa L$.} 
\end{figure}

In Figure 2 we see the corresponding result for $d=3.5$. Here, the 
``double'' crossover between the different fixed points is manifest.   
Asymptotically in the ultraviolet there is a fixed point at $h=1.5$
that corresponds to the $3.5$-dimensional critical point. In the
intermediate region $h$ asymptotes to the critical point of 
the $2.5$-dimensional dimensionally reduced theory while, finally,
in the infrared $h\rightarrow 0$ corresponding to behaviour controlled
by the zero-temperature fixed point that controls the coexistence curve.

Turning now to the anomalous dimension of the field, $\gf$, we find
\be
\gf={2(N-1)\o L\k}t\sum_{n=-\infty}^{\infty}
\int{d^{d-1}y\o(2\pi)^{d-1}}{1\o (y^2+1+{4\pi^2n^2\o L^2\k^2})^2}
\label{gammaphi}
\ee
or in terms of the floating coupling
\be
\gf=\left({N-1\o N-2}\right)h
\label{gffloat}
\ee
\begin{figure}[h]
\centerline{\psfig{figure=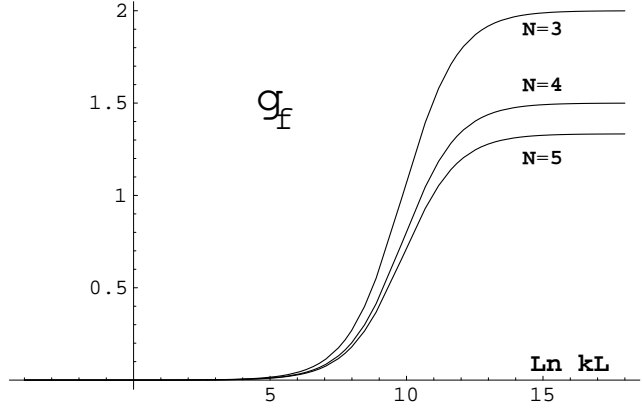,angle=-0,width=3.3in}}
\caption{Graph of $g_f=\gf$ on the separatrix solution of 
(\ref{betah}) as a function of $\ln\kappa L$.} 
\end{figure}
The anomalous dimension also exhibits a dimensional crossover, as can 
be seen in Figure 3, for the case $d=3$, $N=3$, $4$, $5$,
interpolating between the values $\left({N-1\o N-2}\right)(d-2)$ and 
$\left({N-1\o N-2}\right)(d-3)$ in the limits $L\k\ra\infty$ and 
$L\k\ra0$
respectively, where once again we are considering the behaviour near
the critical point. Note that in contrast to the case of an expansion
around the critical point using a $\vf^4$ Landau-Ginzburg-Wilson
Hamiltonian $\gf$ isn't simply the critical exponent $\eta$. This is
due to the fact that the canonical dimension of the fields $\pi$ and
$\sigma$ here is zero. The bulk, $L\rightarrow \infty$, value of 
$\gf$ is $\gf(\infty)=(d-2+\eta)$ where the critical exponent 
$\eta=(d-2)/(N-2)$. In the limit $L\kappa\rightarrow 0$, 
$\kappa\rightarrow\infty$ we see that $\gf\rightarrow (d'-2+\eta')$
where $d'=d-1$ and $\eta'=(d'-2)/(N-2)$ is the critical exponent of
the dimensionally reduced system. In the large-$N$ limit, 
$N\rightarrow\infty$, note that $\gf\rightarrow h=d_{\rm eff}^e$, where
$d_{\rm eff}^e$ is the effective dimension that was found in 
considerations of dimensional crossover in the large-$N$ limit of
a $\lambda\vf^4$ theory \cite{largen}.

\bigskip
\section{Critical Temperature Shift}

The $\beta$ function equation is easily integrated to find
\be
t({\k\o\k_i},L\k)={1\o t_i^{-1}\left({\k_i\o\k}\right)^{d-2}+
\int_{\k_i}^{\k}f(L\k')\left({\k'\o\k}\right)^{d-2}{d\k'\o\k'}}
\label{tsoln}
\ee
where $\k_i$ and $t_i$ are the initial arbitrary renormalization scale and
temperature . The function $f$ is
\be
f(L\k)={2(N-2)\o(4\pi)^{d\o2}}\sum_{n=-\infty}^{\infty}
\int_0^{\infty}{ds\o s^{d-2\o2}}
{\rm e}^{-s}{\rm e}^{-{n^2L^2\k^2\o4s}}
\ee
In the limit $L\ra\infty$ equation (\ref{tsoln}) becomes
\be
t({\k\o\k_i},\infty)={1\o t_i^{-1}\left({\k_i\o\k}\right)^{d-2}+
\int_{\k_i}^{\k}f(\infty)\left({\k'\o\k}\right)^{d-2}{d\k'\o\k'}}
\label{tinfty}
\ee
where $f(\infty)$ simply picks out the $n=0$ term in the sum in $f(L\k)$.
Choosing the initial temperature and initial scale in (\ref{tsoln}) and 
(\ref{tinfty}) to be the same one obtains
\be
\left({1\o T(L)}-{1\o T(\infty)}\right)=\int^{\k}_{\k_i}
(f(L\k')-f(\infty)){\k'}^{d-2}{d\k'\o\k'}\label{temprel}
\ee
where a factor of $\k^{2-d}$ has been absorbed into the dimensionless 
temperature. The interpretation of equation (\ref{temprel}) is that given 
a particular renormalization scale $\k$ in two systems of size $L$ and of 
infinite size then the corresponding temperatures in the two systems are 
related as above. Given that the limit $\k\ra\infty$ corresponds to the approach 
to the critical point we can take this limit in (\ref{temprel}) to find
\be
\left({1\o T_c(L)}-{1\o T_c(\infty)}\right)={1\o2}(f(L\k_i)-f(\infty))
\label{shift1}
\ee
Taking the limit $\k_i\ra0$ corresponds to choosing the initial 
dimensionless temperature to be zero. The shift then becomes
\be
\left({1\o T_c(L)}-{1\o T_c(\infty)}\right)={b_d\o L^{d-2}}\label{shift2}
\ee
where the dimension dependent constant $b_d$ is
\be
b_d={(N-2)\o 2\pi^{d\o2}}\G({d-2\o2})\zeta(d-2)
\ee
The result (\ref{shift2}) is fully in agreement with the expectations of 
finite size scaling \cite{fss} with the exponent $\nu=1/(d-2)$.
In the limit $d\ra3$ $b_d\ra\infty$ as there is a divergence  
in the $\zeta$ function at $d=2$. This corresponds to the fact that the 
the shift is ill defined due to the non-existence of a critical point in
two dimensions as discussed by Barber and Fisher \cite{barbfish}.

\bigskip
\section{Effective Exponents}

A useful set of universal scaling functions are defined by effective 
critical exponents that interpolate between those characteristic 
of the end points of the crossover of interest. Here, given that the 
non-linear $\sigma$-model is restricted to the broken phase we concentrate
on the two effective exponents $\beta_{\rm eff}$ and $\delta_{\rm eff}$
defined as 
\be 
\delta_{\rm eff}^{-1}=\left.{d\ln \sigma\over d\ln H}\right\vert_{t_c(L)}
\qquad 
\beta_{\rm eff}=\left.{d\ln \sigma\over d\ln (t_c(L)-t)}\right\vert_{H=0}
\ee
where $\delta_{\rm eff}$ is defined along the critical isotherm of the
finite size system and $\beta_{\rm eff}$ on the coexistence curve of the 
finite size system. To derive $\beta_{\rm eff}$ we solve the RG equation
for the magnetization, $\bar\vf$ with initial condition $\bar\vf(\kappa=0)=1$.
We then substitute the anomalous dimension (\ref{gffloat}) and consider the
limit $L\kappa\rightarrow\infty$ to find
\be
\beta_{\rm eff}=\beta-(t_c(L)-t){d\over dt}\left({1\over 2}\int_0^{\infty}
\Delta\gf(x,t){dx\over x}\right)
\ee
where $\beta=\nu(d-2+\eta)/2$ is the bulk exponent and 
$\Delta\gf=\gf-(d-2+\eta)$.
 
In the limit $(t_c(L)-t)\rightarrow 0$, $L^{d-2}(t_c(L)-t)\rightarrow\infty$ 
one finds that $\beta_{\rm eff}\rightarrow \nu(d-2+\eta)/2$, 
while in the limit 
$(t_c(L)-t)\rightarrow 0$, $L^{d-2}(t_c(L)-t)\rightarrow 0$ one obtains
$\beta_{\rm eff}\rightarrow \nu'(d'-2+\eta')/2$.
At one-loop order, using equation (\ref{gffloat}), $\beta_{\rm eff}$ 
interpolates between the above asymptotic values, where now 
$\nu=1/(d-2)$, $\eta=(d-2)/(N-2)$, 
$\nu'=1/(d'-2)$, $\eta'=(d'-2)/(N-2)$ and $d'=d-1$.  
%In Figure 3 we see a plot of $\beta_{\rm eff}$ as a function of 
%$\ln(t_c(L)-t)$ the above dimensional crossover being plain to see.

Similarly, one finds for $\delta_{\rm eff}$ that
\be
\delta_{\rm eff}^{-1}={\gf\over (2d-\gf-2\gamma_t)}
\ee
where $\gamma_t\equiv (\beta_t/t)$. In the 
limit $(t_c(L)-t)\rightarrow 0$, $L^{1/\nu}(t_c(L)-t)\rightarrow\infty$ one has
$\gf\rightarrow (d-2+\eta)$ and $\gamma_t\rightarrow 0$. Hence, we see 
that $\delta_{\rm eff}\rightarrow (d+2-\eta)/(d-2+\eta)$.
In the limit $(t_c(L)-t)\rightarrow 0$, 
$L^{1/\nu}(t_c(L)-t)\rightarrow 0$ one finds that $\gf\rightarrow (d'-2+\eta')$
and $\gamma_t\rightarrow 1$. Thus, in this limit 
$\delta_{\rm eff}\rightarrow (d'+2-\eta')/(d'-2+\eta')$, where $d'$ and 
$\eta'$ are as above. At the one-loop level $\gf$ is as given by 
(\ref{gffloat}) and  $\gamma_t=\varepsilon(L\kappa)-h$.
In Figure 4 we see a graph of $\delta_{\rm eff}^{-1}$ as
a function of $\ln L\kappa$ for $d=3$ and $N=3$, $4$, $5$ where the full 
dimensional crossover is evident. 

\begin{figure}[h]
\centerline{\psfig{figure=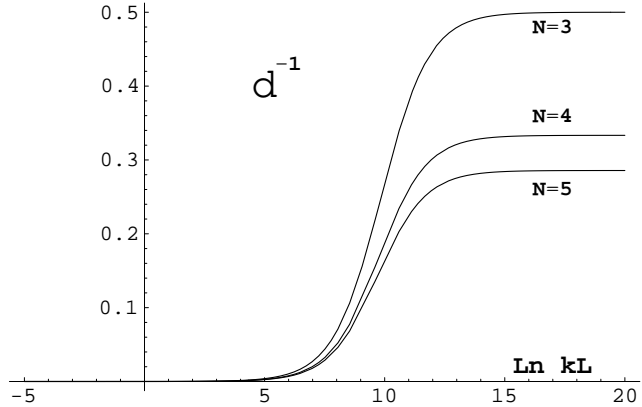,angle=-0,width=3.3in}}
\caption{Graph of $1/d=1/\delta_{\rm eff}$ on the separatrix solution 
of (\ref{betah}) as function of $\ln\kappa L$.} 
\end{figure}

\bigskip
\section{Conclusions}

In this paper we have used environmentally friendly 
renormalization to consider dimensional crossover in the context of a
non-linear $\sigma$-model on a $d$-dimensional film geometry with
periodic boundary conditions. Using an explicitly $L$ dependent
renormalization we derived one loop formulas for the anomalous
dimension of the Goldstone field and for the $\beta$-function,
describing the flow of the temperature as a function of RG scale. We
found that there were three fixed points exhibited in the one
differential equation. A $d$-dimensional critical point, a
$(d-1)$-dimensional critical point and a zero temperature infra-red fixed
point. The $\beta$ function when integrated described the global flow
between all three of these fixed points. Critical temperature shifts
are a particularly interesting consequence of finite size behaviour.
Here, we showed how such shifts could explicitly be calculated finding
an expression to one-loop in agreement with finite size scaling arguments.

Finally, we derived one-loop expressions for the two effective critical
exponents $\beta_{\rm eff}$ and $\delta_{\rm eff}$ showing how they 
interpolated between the asymptotic expressions associated with the 
corresponding $d$ and $(d-1)$-dimensional critical exponents. 
Evidently there is much more that could be done, such as deriving the 
full equation of state etc. Of particular interest will be to adapt the
results to that of a quantum non-linear $\sigma$-model and apply them 
to the case of a high-temperature superconductor. The same mathematical model,
though with a quite different physical interpretation, will also describe
a $d$-dimensional relativistic, quantum field theoretic non-linear 
$\sigma$-model. We hope to return to these interesting issues in a future
publication.

\section{Acknowledgements}

This work was supported by Conacyt grants 30422-E and 32399-E.


\begin{thebibliography}{9}

\bibitem{us}  D. O'Connor and C.R. Stephens,
{\it Nucl. Phys.}\ {\bf B260}, 297 (1991); 
{\it Phys. Rev. Lett.}\ {\bf 72}, 506 (1994);
{\it Int. J. Mod. Phys.}\ {\bf A9}, 2805 (1994);
{\it Physics Reports}, to be published (2002).
\bibitem{sigma} A.A. Migdal, {\it Ah. Eksp. Teor. Fiz.}\ {\bf 69},
810 (1975); A.M. Polyakov, {\it Phys. Lett. }\ {\bf B59}, 79 (1975);
E. Br\'ezin and J. Zinn-Justin, {\it Phys. Rev. Lett.}\ {\bf 36}, 691
(1976).
\bibitem{AG}
D.J. Amit and Y.Y. Goldschmidt, {\it Ann. Phys.}\ {\bf 114}, 356 (1978).
\bibitem{AGsigma}
 D.J. Amit, Y.Y. Goldschmidt and L. Peliti, {\it Ann. Phys.}\ {\bf 116},
1 (1978).
\bibitem{brezzj} E. Br\'ezin and J. Zinn-Justin, {\it Nucl. Phys. }
\ {\bf B257[FS14]}, 867 (1985). 
\bibitem{chak}
S. Chakravarty, B.I. Halperin and D.R. Nelson, 
{\it Phys. Rev. Lett.}\ {\bf 60}, 1057 (1988);
{\it Phys. Rev.}\ {\bf B39}, 2344 (1989).  
\bibitem{largen} D. O'Connor, C.R. Stephens and A.J. Bray, {\it J. Stat. Phys.}
\ {\bf 87}, 273 (1997).
\bibitem{fss}
M.E. Fisher and M.N. Barber, {\it Phys. Rev. Lett.}\ {\bf 28}, 1516 (1972).
\bibitem{barbfish}
M. Barber and M.E. Fisher, {\it Ann. Phys.}\ {\bf 77}, 1 (1973).  
 

\end{thebibliography}
\end{document}